\documentclass[aps,reprint,showpacs,superscriptaddress,
footinbib,citeautoscript]{revtex4-1}%
\usepackage{graphicx}
\usepackage[utf8]{inputenc}
\usepackage{csquotes}
\usepackage[american]{babel}
\usepackage[T1]{fontenc}
\usepackage{enumerate}
\usepackage{mdwlist}
\usepackage[activate=normal]{pdfcprot}
\usepackage{bbding}
\usepackage{color}
\usepackage{amssymb}
\usepackage{amsmath}
\usepackage{amsfonts}
\usepackage{mathrsfs}
\usepackage{bm}
\usepackage{dcolumn}
\usepackage{color}
\usepackage[colorlinks=true,citecolor=blue]{hyperref}%
\providecommand{\U}[1]{\protect\rule{.1in}{.1in}}
\providecommand{\U}[1]{\protect\rule{.1in}{.1in}}
\hypersetup{colorlinks=true,citecolor=blue,linkcolor=red
,urlcolor=blue}

\begin{document}

\title{Indirect coupling of magnons by cavity photons}
\author{Babak Zare Rameshti}
\affiliation{School of Physics, Institute for Research in Fundamental Sciences (IPM), Tehran 19395-5531, Iran}

\author{Gerrit E. W. Bauer}
\affiliation{Kavli Institute of NanoScience, Delft University of Technology, Lorentzweg 1, 2628 CJ Delft, The Netherlands}
\affiliation{Institute for Materials Research and WPI-AIMR, Tohoku University, Sendai 980-8577, Japan}

\begin{abstract}
The interaction between two magnetic spheres in microwave cavities is studied
by Mie scattering theory beyond the magnetostatic and rotating wave
approximations. We demonstrate that two spatially separated dielectric and
magnetic spheres can be strongly coupled over a long distance by standing
cavity modes. The interactions splits acoustical (dark) and optical (bright)
modes in a way that can be mapped on a molecular orbital theory of the
hydrogen molecule. Breaking the symmetry by assigning different radii to the
two spheres introduces \textquotedblleft ionic\textquotedblright\ character to
the magnonic bonds. These results illustrate the coherent and controlled
energy exchange between objects in microwave cavities.

\end{abstract}
\maketitle

\section{Introduction}

\label{sec:intro}

Light-matter systems in which the coherent coupling frequencies exceed the
dissipative loss rates are promising elements for solid state quantum
information circuits~\cite{Wallraff2004, Kubo2010, Putz2014}. Spin ensembles
may couple strongly to electromagnetic modes of a microwave resonator
resulting in hybridized states referred to as magnon
polaritons~\cite{Mills1974, Lehmeyer1985, Cao2015, Zare2015} with the benefit
of long coherence ~\cite{Bar-Gill2013} and short
manipulation~\cite{Childress2006} times. Here a \textquotedblleft
magnon\textquotedblright\ refers to the collective excitation or spin wave of
the polarized spin system. Ferro/ferrimagnets can combine a high spontaneous
spin density with low damping leading to large cooperativities and narrow
linewidths~\cite{Tabuchi2014, Zhang2014}. The strong, and even ultra-strong
coupling regime in which the coupling strength $g$ is comparable to the mode
frequencies~\cite{Niemczyk2010} can therefore be accessed with relative ease.
Furthermore, due to the possibility of coupling magnon modes to photons at
optical frequencies~\cite{Shen1966, Demokritov2001}, magnetic systems are
candidates for coherent conversion of solid state qubits into
\textquotedblleft flying ones\textquotedblright~\cite{Zhang2015, Osada2016}.

On the other hand, controlled creation and read-out of spin-entangled states
in quantum information processing with solid state systems remains a major
challenge. Coherent coupling of spins can be mediated by a variety of physical
mechanisms, such as the magnetic dipolar, exchange, or spin-orbit interaction.
The coupling of spins/pseudospins does not have to be direct, but can be
realized via an intermediary. This can be localized electrons in a filled
shell ion that generate superexchange or the itinerant carriers of metals in
the RKKY interaction~\cite{Ruderman1954, Kasuya1956, Yosida1957}. The
non-local exchange coupling can have either sign; it causes the staggered
magnetization in magnetic multilayers that display the giant
magnetoresistance~\cite{Parkin1990, Bruno1991, Parkin1991}. Quantum systems
can also be coupled radiatively over large distances, i.e. when the
interaction is mediated by virtual photons in a low-loss resonator or
cavity~\cite{Haroche2006, Blais2004}.

Here we address the hybridization of two magnets by cavity photons. Yttrium
iron garnet (YIG), a ferrimagnetic insulator that serves in magnetically
tunable filters and resonators at microwave frequencies, can provide high
coupling strengths and low damping. YIG's spin density is 2$\cdot$10$^{22}$
cm$^{-3}$~\cite{Gilleo1958}, while its Gilbert constant of the magnetization
dynamics typically ranges from 10$^{-3}$ to 10$^{-5}$~\cite{Kajiwara2010,
Heinrich2011, Kurebayashi2011}. Strong coupling between magnons and cavity
photons are manifest in a series of anticrossings in YIG films in coplanar
resonators~\cite{Huebl2013, Stenning2013, Bhoi2014} and YIG spheres in 3D
microwave cavities~\cite{Tabuchi2014, Zhang2014, Goryachev2014}.

Soykal \textit{et al.} \cite{Soykal2010} reported a quantum theory of
photon-magnon coupling in YIG spheres, but this regime has not yet been
reached in experiments. Cao \textit{et al.} modelled the classical
magnon-photon coupling for a thin YIG film in a planar cavity and found strong
coupling even for spin waves beyond the Kittel mode in microwave transmission
and inverse spin Hall effect~\cite{Cao2015}, which was confirmed by
experiments \cite{Bai2015,MaierHaig2016}. Our study of the coherent coupling
between a YIG sphere and microwave cavity modes~\cite{Zare2015} revealed that
YIG\ spheres are efficient antennas for microwaves such that
(ultra)strong-coupling regimes can be achieved in stand-alone magnetic
spheres, as exploited recently~\cite{Bourhill2016}. The long-range strong
coupling of magnons in spatially separated YIG spheres as mediated by a
microwave cavity has been reported~\cite{Zhang2015, Lambert2016}. Electrical
readout of two distant YIG
$\vert$%
Pt bilayers coupled by a microwave cavity mode has been demonstrated
recently~\cite{Bai2017}.

Here we extend the classical model~\cite{Zare2015} to investigate the
long-range coupling of magnons in two spatially separated YIG spheres mediated
by a microwave cavity, producing a delocalized magnon-polariton hybridized
state. The conventional magnetostatic approximation~\cite{Walker1958,
Fletcher1959}, in which the spins interact by the magnetic dipolar field,
disregarding exchange as well as propagation effects, is valid in the Rayleigh
regime $\lambda\gg a$, where $a$ is the radius of the sphere and $\lambda$ the
wavelength of the incident radiation, but breaks down when $\lambda<a$, which
is the regime encountered in sub-mm YIG\ spheres and nanostructured thin
films. We therefore study here the properties of the hybridized
magnon-polaritons, including retardation effects of microwaves, but disregard
the exchange interaction, which is valid for ferromagnets as long as the
exchange length $l_{\mathrm{ex}}=\sqrt{2A/(\mu_{0}M_{s}^{2})}\ll a$, with $A$
and $M_{s}$ being the exchange constant and saturation magnetization,
respectively. Our results help to picture photon-mediated coupling between two
or more magnetic samples in terms of the concept of a chemical bond.

This manuscript is organized as follows. In Sec.~\ref{sec:model}, we introduce
the details of our model and derive the scattered intensity and efficiency
factors for a strongly coupled system of two magnetic spheres in a spherical
microwave cavity. In Sec.~\ref{sec:results}, we present and discuss our
results that demonstrate the effects both due to the dielectric as well as
magnetic effects on the scattering properties and compare our results with
experiments. In Sec.~\ref{sec:concl}, we conclude and summarize our findings.

\section{Model and formalism}

\label{sec:model}

\begin{figure}[ptb]
\includegraphics[width=8.5cm]{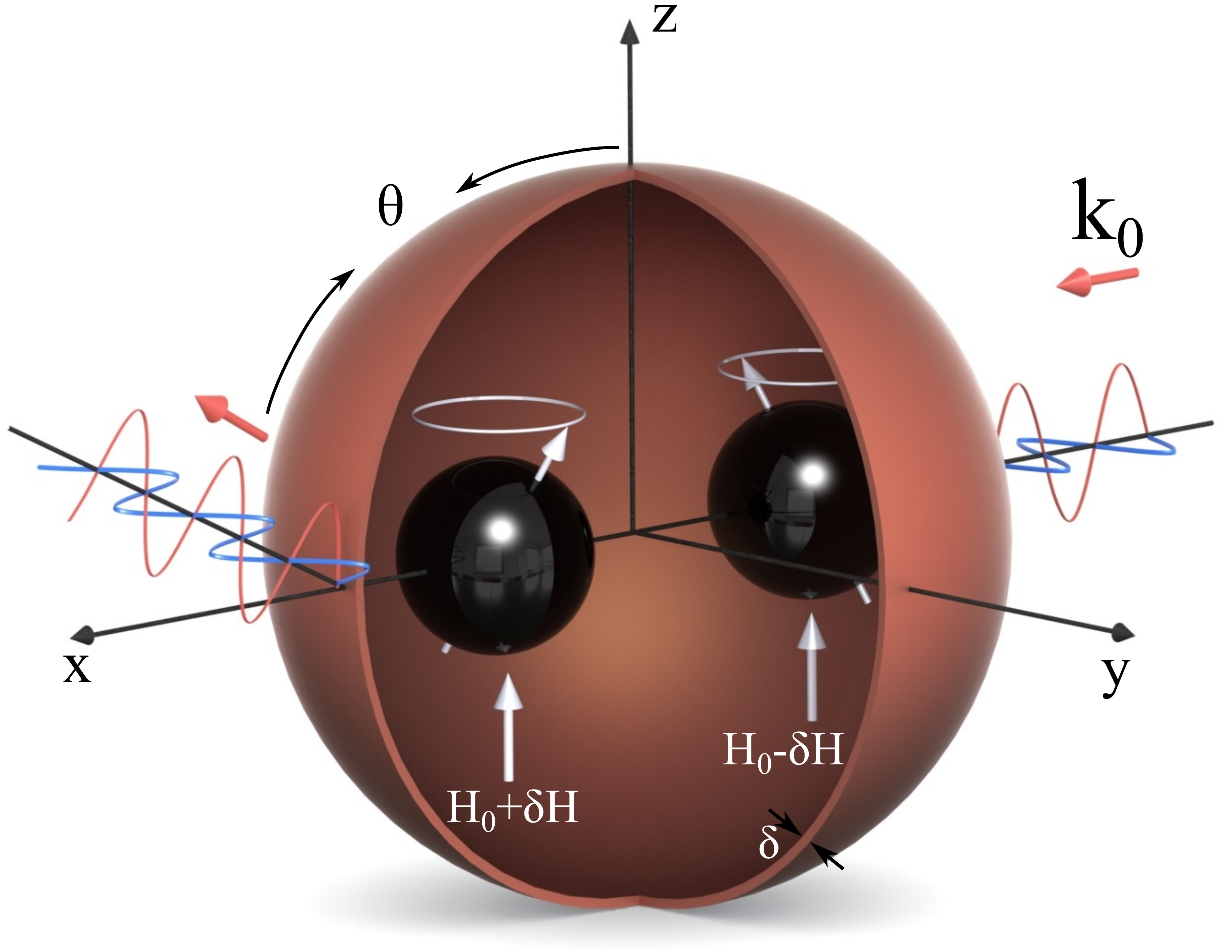}\caption{(Color online) A plane
electromagnetic wave illuminates a large spherical cavity from an arbitray
direction. The latter is modeled by a dielectric spherical shell of radius
$R$, thickness $\delta$, and permittivity $\epsilon_{c}$. Two magnetic spheres
of radius $a_{1}$ and $a_{2}$ are located at antinodes of the ac magnetic
field of the (2,2) and (2,-2) confinement modes of the cavity, i.e., at
$\mathbf{d}_{1}$ and $\mathbf{d}_{2}$ on the $\mathbf{x}$ axis. A constant
magnetic field $H_{0}$ saturates the equilibrium magnetizations. The scattered
waves are measured by a detector in the far field as a function of the
scattering angles, here $\left(  \theta,\varphi\right)  =\left(  \pi
/2,\pi\right)  $.}%
\label{fig1}%
\end{figure}

Mie expressed a general scattering problem in terms of a rapidly converging
expansion into spherical multipole partial waves~\cite{Mie1908, Stratton2007}.
Here we model the indirect coupling of the collective excitations of two
magnetic spheres mediated by photons in a spherical cavity by a Mie-like
expansion of the coupled Landau-Lifshitz-Gilbert and Maxwell equations. We
consider a plane electromagnetic wave with arbitrary polarization and wave
vector shining on a cavity loaded by two magnetic spheres with gyromagnetic
permeability tensors $\overleftrightarrow{\mu}_{1}$ and
$\overleftrightarrow{\mu}_{2}$. A thin spherical shell of a material with high
dielectric constant $\epsilon_{c}/\epsilon_{0}\gg1$, radius $R$, and thickness
$\delta$, models a generic resonant cavity. We mimic realistic situations by
adjusting the parameters $R$ and $\delta$ (see Fig. \ref{fig1}) to tune the
frequencies and broadenings of the cavity modes.

The dynamics of the magnetization vector $\mathbf{M}$ is described by the LLG
equation,
\begin{equation}
\partial_{t}\mathbf{M}=-\gamma\mathbf{M}\times\mathbf{H}_{\mathrm{eff}}%
+\frac{\alpha}{M_{s}}\mathbf{M}\times\partial_{t}\mathbf{M}%
\end{equation}
with $\alpha$ and $M_{s}$ being the damping parameter and saturated
magnetization, respectively. Effective field $\mathbf{H}_{\mathrm{eff}%
}=\mathbf{H}_{\mathrm{ext}}+\mathbf{h}$ comprises the external and (collinear)
easy axis anisotropy fields $\mathbf{H}_{\mathrm{ext}}$ as well as a
distributed ac field $\mathbf{h}(\mathbf{r},t)$. We linearize the LLG equation
by considering the magnetization and driving field vectors %
\begin{align}
\mathbf{M}(\mathbf{r},t)  &  =\mathbf{M}_{0}+\mathbf{m}(\mathbf{r},t)\\
\mathbf{H}(\mathbf{r},t)  &  =\mathbf{H}_{0}+\mathbf{h}(\mathbf{r},t).
\end{align}
To leading order in the small modulations $\mathbf{m}$ and $\mathbf{h}$:
\begin{equation}
\partial_{t}\mathbf{m}=-\gamma(\mathbf{M}_{0}\times\mathbf{H}_{\mathrm{eff}%
}^{(1)}+\mathbf{m}\times\mathbf{H}_{\mathrm{eff}}^{(0)})+\frac{\alpha}{M_{s}%
}\mathbf{M}_{0}\times\partial_{t}\mathbf{m,} \label{eq5}%
\end{equation}
where $\mathbf{H}_{\mathrm{eff}}^{(0)}=\mathbf{H}_{\mathrm{ext}}$ and
$\mathbf{H}_{\mathrm{eff}}^{(1)}=\mathbf{h}$. In the frequency domain, for
$\mathbf{H}_{\mathrm{ext}}$ and $\mathbf{M}_{0}\Vert\hat{z}$,
\begin{equation}
i\omega\mathbf{m}=\mathbf{z}\times\left(  \omega_{\mathrm{M}}\mathbf{h}%
-\omega_{\mathrm{H}}\mathbf{m}+i\omega\alpha\mathbf{m}\right)
\end{equation}
with $\omega_{\mathrm{M}}=\gamma M_{s}$ and $\omega_{\mathrm{H}}=\gamma H_{0}%
$. We express Eq. (\ref{eq5}) as $\mathbf{m}=\overleftrightarrow{\chi}%
\cdot\mathbf{h}$ in terms of the magnetic permeability tensor
\begin{align}
\overleftrightarrow{\mu}  &  =\mu_{0}(\overleftrightarrow{\mathrm{I}%
}+\overleftrightarrow{\chi})\label{eq7}\\
&  =\mu_{0}%
\begin{pmatrix}
1+\chi & -i\kappa & 0\\
i\kappa & 1+\chi & 0\\
0 & 0 & 1
\end{pmatrix}
,
\end{align}
where
\begin{align}
\chi &  =\frac{(\omega_{\mathrm{H}}-i\alpha\omega)\omega_{\mathrm{M}}}%
{(\omega_{\mathrm{H}}-i\alpha\omega)^{2}-\omega^{2}},\\
\kappa &  =\frac{\omega\omega_{\mathrm{M}}}{(\omega_{\mathrm{H}}-i\alpha
\omega)^{2}-\omega^{2}}.
\end{align}

The Maxwell equations inside a homogeneous sphere at frequency $\omega$ read
\begin{align}
\nabla\times\mathbf{E}  &  =i\omega\mathbf{b};\quad\nabla\times\mathbf{h}%
=-i\omega\epsilon_{\mathrm{sp}}\mathbf{E}\\
\nabla\cdot\mathbf{E}  &  =0;\quad\nabla\cdot\mathbf{b}=0. \label{eq10}%
\end{align}
The magnetic induction $\mathbf{b}$ and the magnetic field $\mathbf{h}$ inside
this medium are related by
\begin{equation}
\mathbf{b}=\overleftrightarrow{\mu}\cdot\mathbf{h},\quad\mathbf{D}%
=\epsilon_{\mathrm{sp}}\mathbf{E}.
\end{equation}
and $\mathbf{b}$ satisfies the wave equation
\begin{equation}
\nabla\times\nabla\times\left(  \mu_{0}\overleftrightarrow{\mu}^{-1}%
\cdot\mathbf{b}\right)  -k_{\mathrm{sp}}^{2}\mathbf{b}=0, \label{wave_eq}%
\end{equation}
where $k_{\mathrm{sp}}^{2}=\omega^{2}\epsilon_{\mathrm{sp}}\mu_{0}$ and
$\epsilon_{\mathrm{sp}}$ is the scalar permittivity of the medium. Keeping Eq.
(\ref{eq10}) in mind, we expand $\mathbf{h}$ in terms of vector spherical
waves as
\begin{equation}
\mathbf{h}=\sum_{nm}\bar{\eta}_{nm}\left[  d_{mn}\mathbf{V}_{nm}%
^{(1)}(k,\mathbf{r})+c_{mn}\mathbf{N}_{nm}^{(1)}(k,\mathbf{r})\right]
\label{eq16}%
\end{equation}
where $k$ is as yet undetermined, $n$ runs from $1$ to $\infty,$ and
$m=-n,\cdots,n.$ The prefactors read $\bar{\eta}_{nm}=\eta_{nm}k_{0}%
/(\omega\mu_{0})$ with
\begin{equation}
\eta_{nm}=i^{n}E_{0}\left[  \frac{2n+1}{n(n+1)}\frac{(n-m)!}{(n+m)!}\right]
^{1/2}%
\end{equation}
where $E_{0}$ is the amplitude of the electric field of the incident wave. The
vector spherical wave functions are defined as
\begin{align}
\mathbf{V}_{nm}^{(j)}(k,\mathbf{r})  &  =z_{n}^{(j)}(kr)\mathbf{X}%
_{nm}(\mathbf{r}),\nonumber\\
k\mathbf{N}_{nm}^{(j)}(k,\mathbf{r})  &  =\nabla\times\mathbf{V}_{nm}%
^{(j)}(k,\mathbf{r}).
\end{align}
where $z_{n}^{(j)}$ are spherical Bessel functions of the $j$-th kind, e.g.,
$z_{n}^{(3)}=h_{n}^{(1)}$ is the spherical Bessel functions of the third kind
(Hankel function). $\mathbf{X}_{nm}=\mathbf{L}Y_{nm}(\hat{r})/\sqrt{n(n+1)}$,
where $Y_{nm}(\hat{r})$ are spherical (surface) harmonics and $\mathbf{L}%
=-i\mathbf{r}\times\nabla_{r}$ is the angular momentum and $\nabla_{r}$ the
gradient operator. By invoking the vector spherical wave function expansion
for $\mathbf{b}$ and $\overleftrightarrow{\mu}^{-1}\cdot\mathbf{b}$ in the
wave equation Eq. (\ref{wave_eq}) leads to the dispersion relation for
$k(\omega)$. We focus in the following on the lowest frequency resonances for
a given angular momentum without radial nodes in the sphere. For simplicity of
notation we therefore omit the \textquotedblleft main quantum
number\textquotedblright\ when labelling the cavity modes.

The electric field distribution is obtained by $\mathbf{E}=(i/\omega
c)\nabla\times\mathbf{h}$. We expand the incident fields $\mathbf{E}%
_{\mathrm{inc}}$, $\mathbf{h}_{\mathrm{inc}}$ and scattered fields
$\mathbf{E}_{s}$, $\mathbf{h}_{s}$ outside the sphere analogously. The
scattered field reads then
\begin{equation}
\mathbf{h}_{s}=\sum_{nm}\bar{\eta}_{nm}\left[  b_{mn}\mathbf{N}_{nm}%
^{(3)}+a_{mn}\mathbf{V}_{nm}^{(3)}\right]
\end{equation}
with $k_{0}^{2}=\omega^{2}\epsilon_{0}\mu_{0}$. The expansion coefficients
$a_{nm}$ and $b_{nm}$ are determined by the boundary conditions. We consider
the situation that the magnetic sphere is illuminated by a plane wave with
arbitrary polarization and direction of incidence as indicated in Fig.
(\ref{fig1}). This incident fields can be expanded as,
\begin{equation}
\mathbf{h}_{\mathrm{inc}}=-\sum_{nm}\bar{\eta}_{nm}\left[  q_{mn}%
\mathbf{N}_{nm}^{(1)}+p_{mn}\mathbf{V}_{nm}^{(1)}\right]
\end{equation}
with coefficients
\begin{align}
p_{mn}  &  =\frac{\eta_{nm}}{i^{n}E_{0}}\left[  p_{\theta}\tau_{mn}(\cos
\theta_{k})-ip_{\phi}\pi_{mn}(\cos\theta_{k})\right]  e^{-im\phi_{k}%
}\label{eq54}\\
q_{mn}  &  =\frac{\eta_{nm}}{i^{n}E_{0}}\left[  p_{\theta}\pi_{mn}(\cos
\theta_{k})-ip_{\phi}\tau_{mn}(\cos\theta_{k})\right]  e^{-im\phi_{k}}
\label{eq55}%
\end{align}
where $\hat{\mathbf{p}}=(p_{\theta}\boldsymbol{\hat{\theta}}_{k}+p_{\phi
}\boldsymbol{\hat{\phi}}_{k})$ is the normalized complex polarization vector,
with unit vectors $\boldsymbol{\hat{\theta}}_{k}$ and $\boldsymbol{\hat{\phi}%
}_{k},$ $\left\vert \hat{\mathbf{p}}\right\vert =1$ and $\theta_{k}(\phi_{k})$
is the polar (azimuthal) angle of incidence. Two auxiliary functions are
defined by
\begin{equation}
\pi_{mn}(\cos\theta)=\frac{m}{\sin\theta}P_{n}^{m}(\cos\theta),~\tau_{mn}%
(\cos\theta)=\frac{d}{d\theta}P_{n}^{m}(\cos\theta)
\end{equation}
All fields of the scattering problem are now expanded in terms of vector
spherical wave functions. The boundary conditions
\begin{align}
\left[  \mathbf{E}_{\mathrm{inc}}+\mathbf{E}_{s}\right]  \times\mathbf{e_{r}}
&  =\mathbf{E}_{i}\times\mathbf{e_{r},}\\
\left[  \mathbf{h}_{\mathrm{inc}}+\mathbf{h}_{s}\right]  \times\mathbf{e_{r}}
&  =\mathbf{h}_{i}\times\mathbf{e_{r}}%
\end{align}
can be rewritten in terms of the transmission matrix $\mathcal{T}$ that
relates the scattered to the incoming fields
\begin{equation}%
\begin{pmatrix}
a_{nm}\\
b_{nm}%
\end{pmatrix}
=\mathcal{T}%
\begin{pmatrix}
p_{nm}\\
q_{nm}%
\end{pmatrix}
. \label{sscatcoeff}%
\end{equation}
We are interested in more than one scattering objects in the cavity. In order
to describe the collective excitations of non-overlapping magnetic spheres, we
expand the total incident field striking the surface of the $i$-th sphere, the
initial incident waves, and the scattered field of the other spheres with
index $j\neq i$, in the coordinate systems centered at sphere $i$ as
\begin{equation}
\mathbf{E}_{\mathrm{inc}}^{i}=\mathbf{E}_{\mathrm{inc}}+\sum_{j\neq
i}\mathbf{E}_{s}^{j};\quad\mathbf{h}_{\mathrm{inc}}^{i}=\mathbf{h}%
_{\mathrm{inc}}+\sum_{j\neq i}\mathbf{h}_{s}^{j}%
\end{equation}
The transformation of waves scattered by one sphere into incident waves for
the other one is formulated by the addition theorem of vector spherical
harmonics~\cite{Xu1996}, i.e., the expansion of the basis set in a translated
reference system. By transforming the wave scattered by one sphere to a
coordinate system centered at the other and imposing appropriate boundary
conditions, we arrive at the scattering coefficients
\begin{equation}%
\begin{pmatrix}
a_{nm}^{i}\\
b_{nm}^{i}%
\end{pmatrix}
=\mathcal{T}^{i}\left[
\begin{pmatrix}
p_{nm}^{i}\\
q_{nm}^{i}%
\end{pmatrix}
+\sum_{j\neq i}\mathcal{R}^{ji}%
\begin{pmatrix}
a_{nm}^{j}\\
b_{nm}^{j}%
\end{pmatrix}
\right]  ,
\end{equation}
where the superscript indicates the coordinate system centered at sphere $i$
and $\mathcal{R}^{ji}$ is the translation matrix from sphere $j$ to
$i$~\cite{Xu1996}. The second term on the right-hand side represents the
multiple scattering between the objects. The scattering coefficients in the
coordinate system of the cavity can be obtained by the unitary transformation
$\mathcal{R}^{i0}$ defined by the addition theorem
\begin{equation}%
\begin{pmatrix}
a_{nm}^{0}\\
b_{nm}^{0}%
\end{pmatrix}
=\mathcal{R}^{i0}%
\begin{pmatrix}
a_{nm}^{i}\\
b_{nm}^{i}%
\end{pmatrix}
\end{equation}
These expressions are sufficient to compute the scattering matrix for the
entire system.

In order to make contact with experiments, we consider the \textit{far-field}
limit, in which the intensity of the two polarization components $I_{\theta}$
and $I_{\phi}$ are
\begin{equation}
I_{\theta}\sim\frac{E_{0}^{2}}{k_{0}^{2}r^{2}}|S_{1}(\theta,\phi)|^{2},\qquad
I_{\phi}\sim\frac{E_{0}^{2}}{k_{0}^{2}r^{2}}|S_{2}(\theta,\phi)|^{2}%
\end{equation}
where $\theta(\phi)$ is the polar (azimuthal) angle of the observer at
distance $r$ and scattering intensity functions are
\begin{align}
S_{1}(\theta,\phi)  &  =\sum_{nm}\left[  a_{mn}\tilde{\tau}_{mn}\left(
\cos\theta\right)  +b_{mn}\tilde{\pi}_{mn}\left(  \cos\theta\right)  \right]
e^{im\phi},\qquad\\
S_{2}(\theta,\phi)  &  =\sum_{nm}\left[  a_{mn}\tilde{\pi}_{mn}\left(
\cos\theta\right)  +b_{mn}\tilde{\tau}_{mn}\left(  \cos\theta\right)  \right]
e^{im\phi},\qquad
\end{align}
We define a dimensionless \textit{scattering efficiency factor }%
$Q_{\mathrm{sca}}$ as the total (i.e. angular integrated) scattering cross
section of the light intensity divided by the geometrical area $\pi R^{2}$
as,
\begin{equation}
Q_{\mathrm{sca}}=\dfrac{4}{k_{0}^{2}R^{2}}\sum_{nm}\left(  |a_{nm}%
|^{2}+|b_{nm}|^{2}\right)
\end{equation}
The efficiency factor $Q_{\mathrm{ext}}$ defined analogously for the total
extinction cross section
\begin{equation}
Q_{\mathrm{ext}}=\dfrac{4}{k_{0}^{2}R^{2}}\sum_{nm}\operatorname{Re}\left(
p_{nm}^{\ast}a_{nm}+q_{nm}^{\ast}b_{nm}\right)
\end{equation}
measures the total energy loss of the incident beam by absorption and
scattering.
\begin{equation}
Q_{\mathrm{abs}}=Q_{\mathrm{ext}}-Q_{\mathrm{sca}} \label{Qsca}%
\end{equation}
reflects the loss of intensity due to Gilbert damping in the sample.

\section{Results}

\label{sec:results}

The observables defined above can be computed numerically as a function of
material and cavity parameters. We focus here on a spherical cavity with fixed
radius ($R=4$\thinspace mm) loaded with two dielectric spheres at a fixed
distance $d_{0}=2.5$\thinspace mm, but with adjustable diameter, as in Fig.
\ref{fig1}. We focus on the strong coupling regime in which the polaritonic
mode splitting is comparable or larger than the dissipation, i.e. we have
spectrally sharp cavity modes and not too large Gilbert damping. Without using
the macrospin approximation, we focus our discussion to the nearly uniform
(Kittel) mode that displays the strongest coupling to the
microwaves~\cite{Cao2015}.

Forward scattered intensities, i.e., $\theta=\pi/2,\phi=\pi$, and scattering
efficiency factors are convenient and observable measures of the
microwave-matter coupling. In order to compare our results with recent
experiments, we adopt parameters for YIG with gyromagnetic ratio $\gamma
/(2\pi)=28$ GHz/T, saturation magnetization $\mu_{0}M_{s}=175$ mT~
\cite{Manuilov2009}, Gilbert damping constant $\alpha=3\times10^{-4}%
$~\cite{Kajiwara2010, Heinrich2011, Kurebayashi2011}, and relative
permittivity $\epsilon_{\mathrm{sp}}/\epsilon_{0}=15$~\cite{Sadhana2009}. The
incident microwave radiation comes from the positive $\mathbf{x}$ direction
($\theta_{k}=\pi/2$ and $\phi_{k}=0$) and is linearly polarized such that its
electric/magnetic components are in the $-\mathbf{z}/\mathbf{y}$ directions
(static magnetic field and magnetization $\mathbf{H}_{0}\parallel\mathbf{z}$).
We also investigate the dependence of the observables on the scattering angle
with respect to the outgoing radiation. \begin{figure}[ptb]
\includegraphics[width=8.5cm]{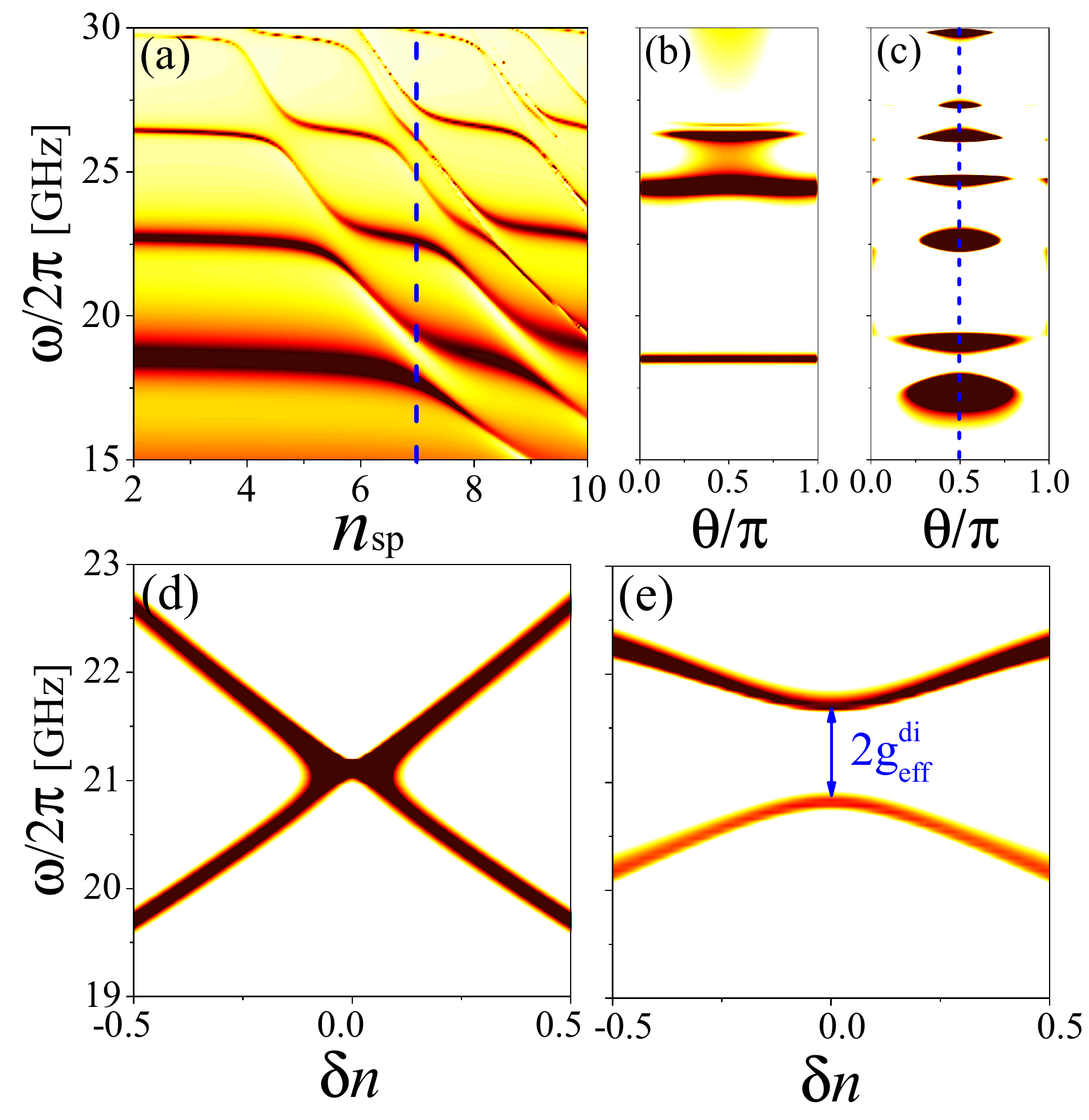}\caption{(Color online) (a): The scattering
efficiency factor Eq. (\ref{Qsca}) for two non-magnetic dielectric spheres of
radius $a_{1}=a_{2}=1$ mm, cavity radius $R=4$ mm, and asymmetry $\delta n=1$
plotted as a function of frequency $\omega/2\pi$ and average refractive index
$n_{\mathrm{sp}}$. (b) and (c): The scattering intensity $|S_{1}|^{2}$ as
function of scattering angle $\theta$ and frequency $\omega/2\pi$ plotted for
the same spheres ($n_{\mathrm{sp}}=7,$ $\delta n=1$) without and with cavity,
respectively, while (d) and (e) are the corresponding scattering efficiencies.
The anticrossing in (e) reveals the interaction with the cavity field by the
coupling strength $2g_{\mathrm{eff}}^{\mathrm{di}}$, i.e., the frequency
splitting of the modes at $\delta n=0$. The dashed lines are guides for the
eye.}%
\label{fig2}%
\end{figure}

We start by studying the effects of asymmetry on the photon-mediated coupling
of two \emph{non-magnetic} spheres with refractive indices $n_{1}%
=n_{\mathrm{sp}}+\delta n$ and $n_{2}=n_{\mathrm{sp}}-\delta n$. In Fig.
\ref{fig2}(a) the scattering efficiency factor Eq. (\ref{Qsca}) is plotted as
a function of frequency $\omega/\left(  2\pi\right)  $ and average refractive
index $n_{\mathrm{sp}}=\sqrt{\epsilon_{\mathrm{sp}}/\epsilon_{0}}$ of the
spheres with $a=1$ mm in a spherical cavity with radius $R=4$ mm and broken
symmetry with $\delta n=1$. The spheres are placed at the local maxima of the
electric field distribution of the cavity, i.e., $\mathbf{d}_{1}%
=d_{0}\mathbf{x}$ and $\mathbf{d}_{2}=-d_{0}\mathbf{x}$, respectively, where
$d_{0}=2.5$\thinspace mm. This ensures a significant coupling strength and
nearly uniform distribution of the cavity field over the spheres.

When $\delta n\neq0$ the individual resonances of the two spheres are
distinguishable in Fig. \ref{fig2}(a). Not only the lowest but also higher
plasmonic modes ($\sim n_{\mathrm{sp}}^{2})$ anticross strongly with the
(constant) cavity resonances. The angular dependence of the scattering without
and with cavity is plotted in panels (b) and (c) of Fig. \ref{fig2},
respectively. The eigenmodes of the two coupled-dielectric spheres have a
predominant $s$-wave character when the wavelength $\lambda\gtrapprox
a\sqrt{\epsilon_{\mathrm{sp}}/\epsilon_{0}}$, i.e. no scattering-angle
dependence in the regime in which no resonant states are formed.

The radiative coupling between two dielectric spheres by the cavity eigenmodes
is revealed by tuning the resonances with the asymmetry parameter $\delta
n\in\lbrack-0.5,0.5]$ for $n_{\mathrm{sp}}=7$. Fig. \ref{fig2}(d) and (e) are
plots of the scattering efficiency factor $Q_{\mathrm{sca}}$ as a function of
frequency $\omega/2\pi$ and asymmetry $\delta n$ in the absence and presence
of the external cavity, respectively. The photon-mediated coupling corresponds
to the splitting at the nominal crossing point ($\delta n=0$) and found to be
$g_{\mathrm{eff}}^{\mathrm{di}}/2\pi\sim0.6$ GHz, which is much larger that
the broadening and therefore \textquotedblleft strong\textquotedblright.
Removing the cavity suppresses the splitting, as seen in Fig. \ref{fig2}(d),
proving that the direct dipolar coupling between the spheres and the multiple
scattering of the microwaves between spheres in the absence of a cavity are
weak. In analogy with plasmonic molecules in metallic
nanostructures~\cite{Prodan}, which are bound by the optical near-fields, we
refer to this hybridized state as a \emph{plasmon-polariton molecule}.
\begin{figure}[ptb]
\includegraphics[width=8.5cm]{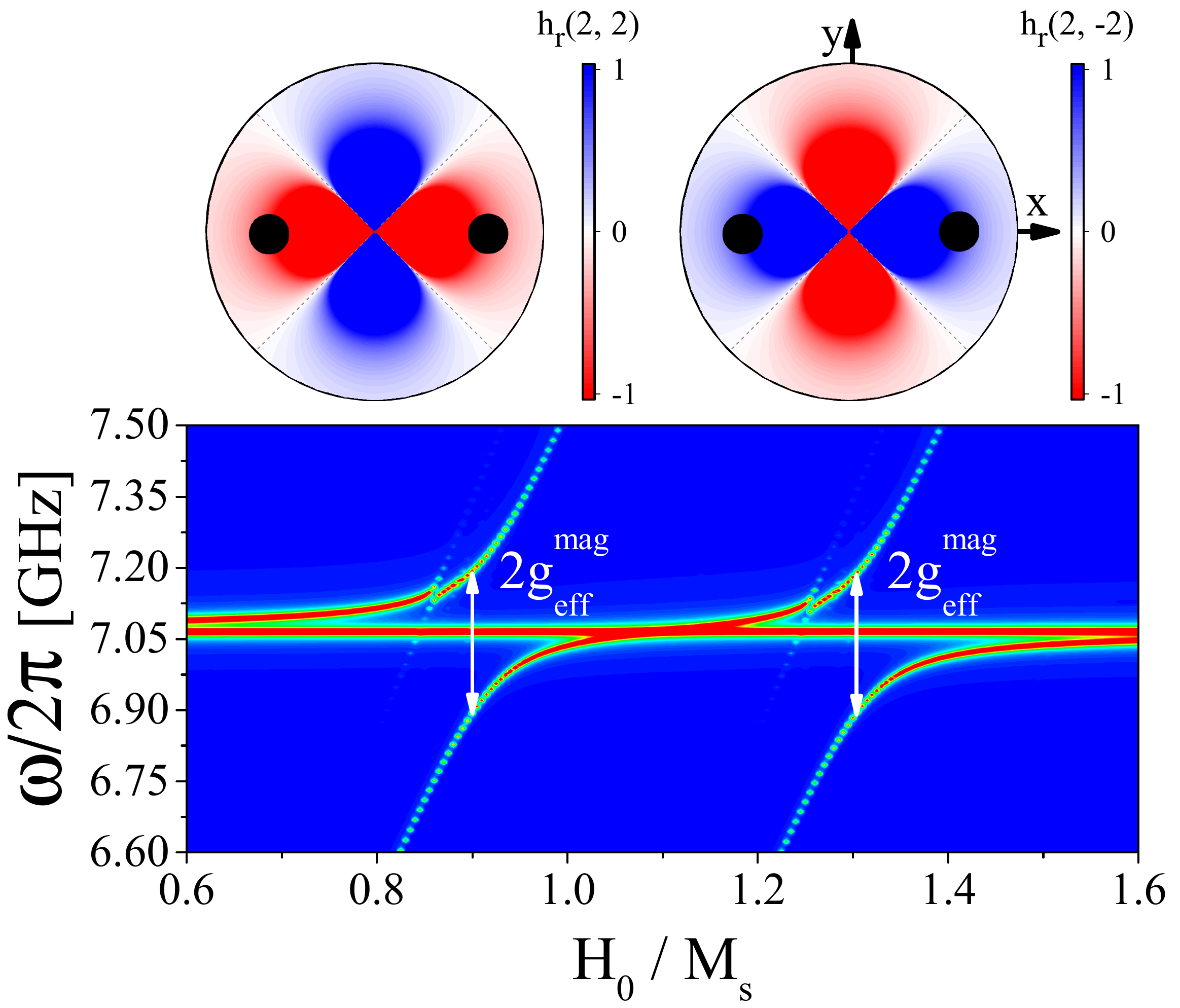}\caption{(Color online) Scattering
efficiency factor $Q_{\mathrm{sca}}$ as function of magnetic field
$H_{0}/M_{s}$ and frequency $\omega/2\pi$ for two YIG spheres of radius
$a_{1}=a_{2}=0.5$ mm and relative permittivity $\epsilon_{\mathrm{sp}%
}/\epsilon_{0}=15$ in a spherical cavity of radius $R=4$ mm on the two
antinodes of the cavity mode at $\omega_{2}/2\pi\sim7.05$ GHz is shown in
bottom panel. The field at each sphere is detuned by $\left\vert \delta
H\right\vert /M_{s}\sim0.2$ with opposite sign. $g_{\mathrm{eff}%
}^{\mathrm{mag}}$ is the magnon-cavity coupling strength. The radial component
of microwave magnetic field $h_{r}$ for the cavity mode frequency $\omega_{2}$
in the equator plane is shown in top panel, the black circles indicate two
spheres.}%
\label{fig3}%
\end{figure}

\begin{figure}[ptb]
\includegraphics[width=8.5cm]{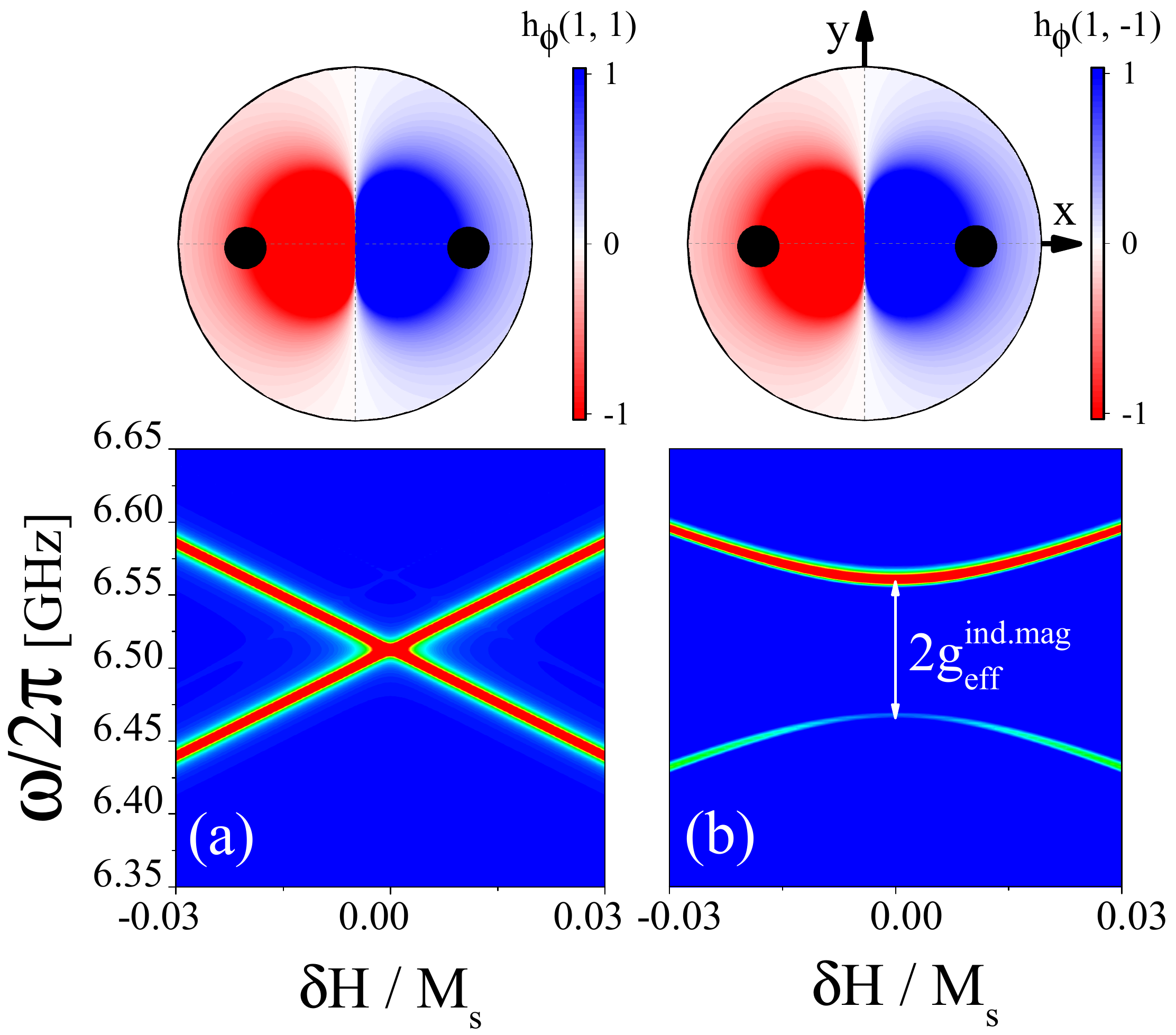}\caption{(Color online) (a) and (b):
Scattering efficiency factor $Q_{\mathrm{sca}}$ as function of $\omega/2\pi$
and $\delta H/M_{s}$ for the same two spheres as Fig. (\ref{fig3}) without and
with cavity, respectively, but the detuning is much smaller than in Fig.
(\ref{fig3}). $H_{0}/M_{s}=1$ is fixed such that the magnetostatic modes of
each sphere are detuned from $\omega_{1}$. The anticrossing in (b) illustrates
the coupling of the two YIG spheres; the non-local magnon-magnon coupling
strength $g_{\mathrm{eff}}^{\mathrm{ind.mag}}$ is the frequency splitting of
the modes at $\delta H=0$. The azimuthal component of microwave magnetic field
$h_{\phi}$ for the cavity mode frequency $\omega_{1}$ in the equator plane is
shown in top panel, the black circles indicate two spheres.}%
\label{fig4}%
\end{figure}

The magnetism of the spheres affects the microwave scattering properties
strongly, but the plasmonic effects causing hybridization of the resonances of
cavity and sphere remain to be very relevant. Our results help to interpret
recent experimental results on cavity-mediated coupling of two YIG
spheres~\cite{Lambert2016} by taking into acount the finite size of the
spheres and cavity-field distribution. Fig. \ref{fig3} shows the scattering
efficiency factor as a function of frequency $\omega/2\pi$ and uniform
magnetic field $H_{0}/M_{s}$ for our spherical cavity containing now two YIG
spheres with radii $a_{1}=a_{2}=0.5$ mm. The frequency of the microwaves with
wave vector along the $x$-direction is tuned to the 5-fold degenerate cavity
modes with $n=2$ ($d$-wave); $\omega_{2}/2\pi\sim7.05$ GHz, of which only the
$\omega_{2,\pm2}$ states are excited by symmetry. An asymmetry is now induced
by a detuning magnetic field with opposite sign on different spheres $\delta
H=\pm0.2\,M_{s}$. The two spheres occupy antinodes of the $p$ and $d$ cavity
resonances shown in the top panels of Figs. \ref{fig3} and \ref{fig4} with
parameters chosen to be close to the experiment~\cite{Lambert2016}. Two
distinct anticrossings are the signature of mixed \textit{magnon-polariton}
modes with a magnon-photon coupling of $g_{\mathrm{eff}}^{\mathrm{mag}}%
/2\pi\sim150$ MHz between the Kittel modes of both spheres and the cavity
mode. Small satellites indicate the coupling to a higher (\textquotedblleft
Walker\textquotedblright) mode in both spheres. Fig. \ref{fig3} also shows a
cavity mode that is not affected by the magnets~\cite{Zhang2015}. This mode is
a linear combination of the active cavity modes $\omega_{2,\pm2}$ that does
not couple to the sphere. Although the spherical symmetry of the empty cavity
has been broken by the load, the axial symmetry remains intact and is
responsible for this effect. \begin{figure}[ptb]
\includegraphics[width=8.5cm]{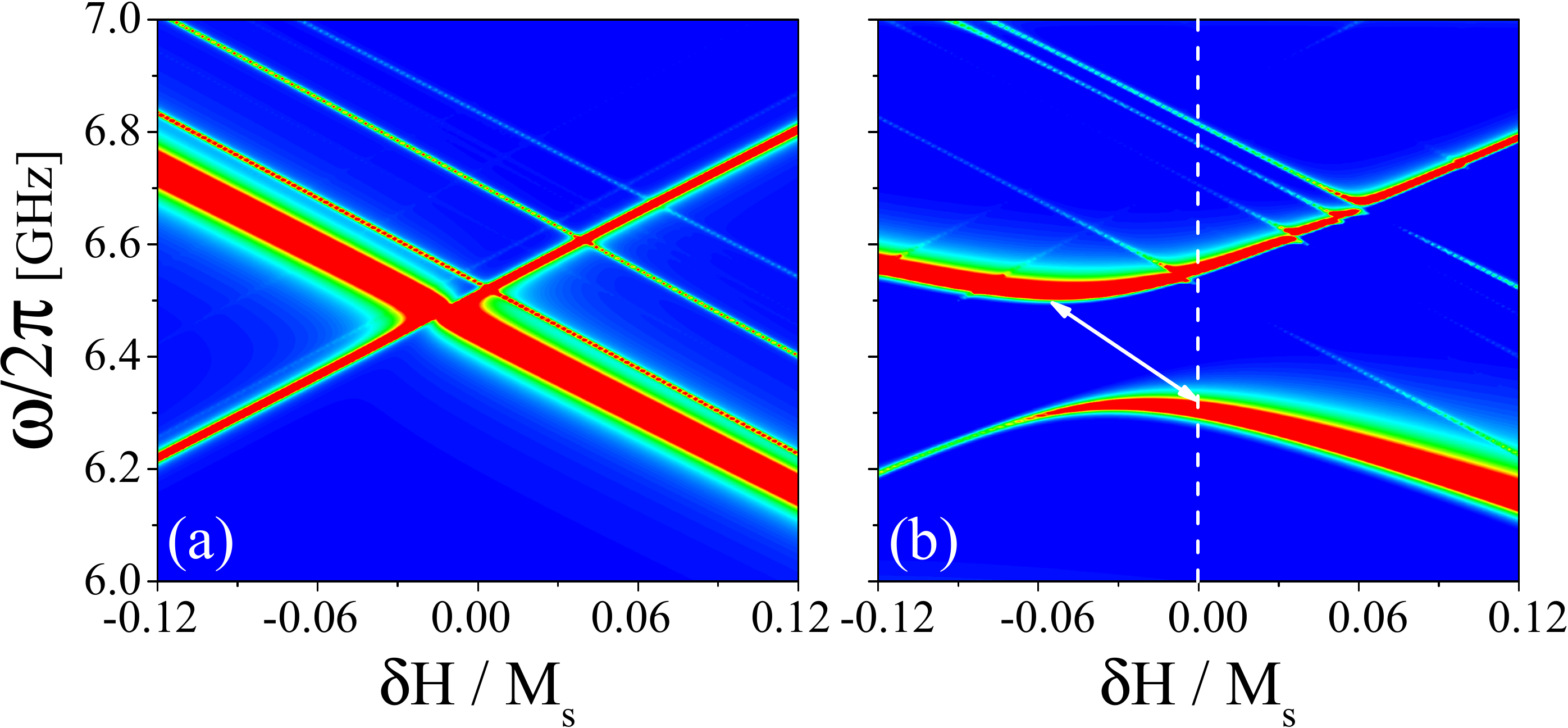}\caption{(Color online) Scattering
efficiency as function of frequency $\omega/2\pi$ and normalized bias field
$\delta H/M_{s}$ for two YIG spheres ($\epsilon_{\mathrm{sp}}/\epsilon_{0}%
=15$) with radius $a_{1}=0.5$ mm and $a_{2}=1$ mm (a) without cavity and (b)
in a spherical cavity of radius $R=4$ mm. The white arrow in (b) illustrates
the \textquotedblleft indirect gap\textquotedblright\ induced by the radiative
coupling.}%
\label{fig5}%
\end{figure}

Next we fix $H_{0}/M_{s}=1$ and study the effect of small detunings $\delta
H/M_{s}$ in the dispersive regime$.$ In Figs. \ref{fig4}(a) and (b) the Kittel
mode lies above the $p$-wave cavity eigenmode $\omega_{1}/2\pi\sim6\,$GHz.
Note that the scattering efficiencies in the dispersive regime are much
smaller than those in Fig. \ref{fig3}. Panel (a) shows results for two YIG
spheres of radii $a=0.5$ mm without cavity, while for panel (b) the spherical
cavity has been added. The anticrossing in Fig. \ref{fig4}(b) illustrates that
the magnons of the two magnets interact over long distances through the
virtual exchange of cavity microwave photons. The coupling strength is given
by the frequency splitting of the modes at $\delta H=0$, giving a value of
$g_{\mathrm{eff}}^{\mathrm{ind.mag}}/2\pi\sim43$ MHz. This coupling requires
an external resonator, cf. Fig. \ref{fig4}(a), and can therefore not be
explained by the direct magnetic dipolar interactions or multiple scattering
between the spheres, as observed \cite{Lambert2016}.

We observe that the upper mode has a relatively large oscillator strength
(\textquotedblleft bright mode \textquotedblright), while the lower mode
intensity is suppressed at $\delta H=0$ (\textquotedblleft dark
mode\textquotedblright). The order and symmetry of these modes depends on the
sign of the magnon-cavity mode detuning as well as the phase relation between
the amplitude of the cavity mode on the spheres. In principle, many modes
contribute, but the ones closest in frequency dominate. The higher frequency
mode in \ref{fig4}(a) is the \textquotedblleft acoustic\textquotedblright%
\ (symmetric) mode that strongly interacts with the low frequency mode
$\omega_{1}$, which has the largest oscillator strength for forward
scattering. The lower \textquotedblleft optical\textquotedblright%
\ (antisymmetric) mode for $\delta H=0$ interacts with (and is pushed to lower
frequencies) by mode $\omega_{2}$. The scattering power of the $\omega_{2}$
mode (without load) is much weaker than that of $\omega_{1},$ which renders
the lower collective magnetic mode to be \textquotedblleft
dark\textquotedblright. We note that the \textquotedblleft
darkness\textquotedblright\ is not absolute, since the remaining intensity
does not vanish for $\delta H=0$ and depends on the details of the system and
scattering configuration.

Lambert et al. \cite{Lambert2016} find that a cavity mode $\omega_{2}/2\pi
\sim7.15$ GHz couples with the Kittel mode of a YIG sphere with $a=0.5$%
\thinspace mm by $g_{2}/2\pi\left(  \equiv g_{\mathrm{eff}}^{\mathrm{mag}%
}/2\pi\right)  \approx150$ MHz, in excellent agreement with our calculations.
By a dispersive measurement technique they also observe a splitting which they
interpret in terms of in-phase and out-of-phase precessions of the individual
magnetization dynamics. The observed splitting of these two modes agrees well
with the calculated ones, i.e. $2J/2\pi=87~$MHz as compared to our
$2g_{\mathrm{eff}}^{\mathrm{ind.mag}}/2\pi\sim86~$MHz. The order of
\textquotedblleft bright\textquotedblright\ and \textquotedblleft
dark\textquotedblright\ modes is opposite to what we find in Fig. \ref{fig4}.
This discrepancy is caused by the relative low frequency $\omega_{1}/2\pi
\sim3.55\,$GHz in the experiments, which is not reproduced by our spherical
cavity in which $\omega_{1}/2\pi\sim6$ GHz.

For two identical spheres the scattering properties $\mathcal{A}$, such as
$Q_{\mathrm{sca}}$, are parity (mirror) symmetric in parameter space, i.e.,
$\mathcal{A}(\omega,\delta H)=\mathcal{A}(\omega,-\delta H)$. The mode
coupling at $\delta H=0$ therefore must generate a direct gap and parabolic
dependence on small $\delta H,$ as indicated in Fig. \ref{fig4}(b). Different
radii break the symmetry and $\mathcal{A}(\omega,\delta H)\neq\mathcal{A}%
(\omega,-\delta H)$. Fig. \ref{fig5}(b) illustrates the strong magnon-magnon
coupling of two different YIG spheres with $a_{1}=1$ mm and $a_{2}=0.5$ mm
$\left(  \epsilon_{\mathrm{sp}}/\epsilon_{0}=15\right)  $ in a cavity of
radius $R=4$ mm. The asymmetry generates now an \textquotedblleft
indirect\textquotedblright\ gap.

\begin{figure}[ptb]
\includegraphics[width=8.5cm]{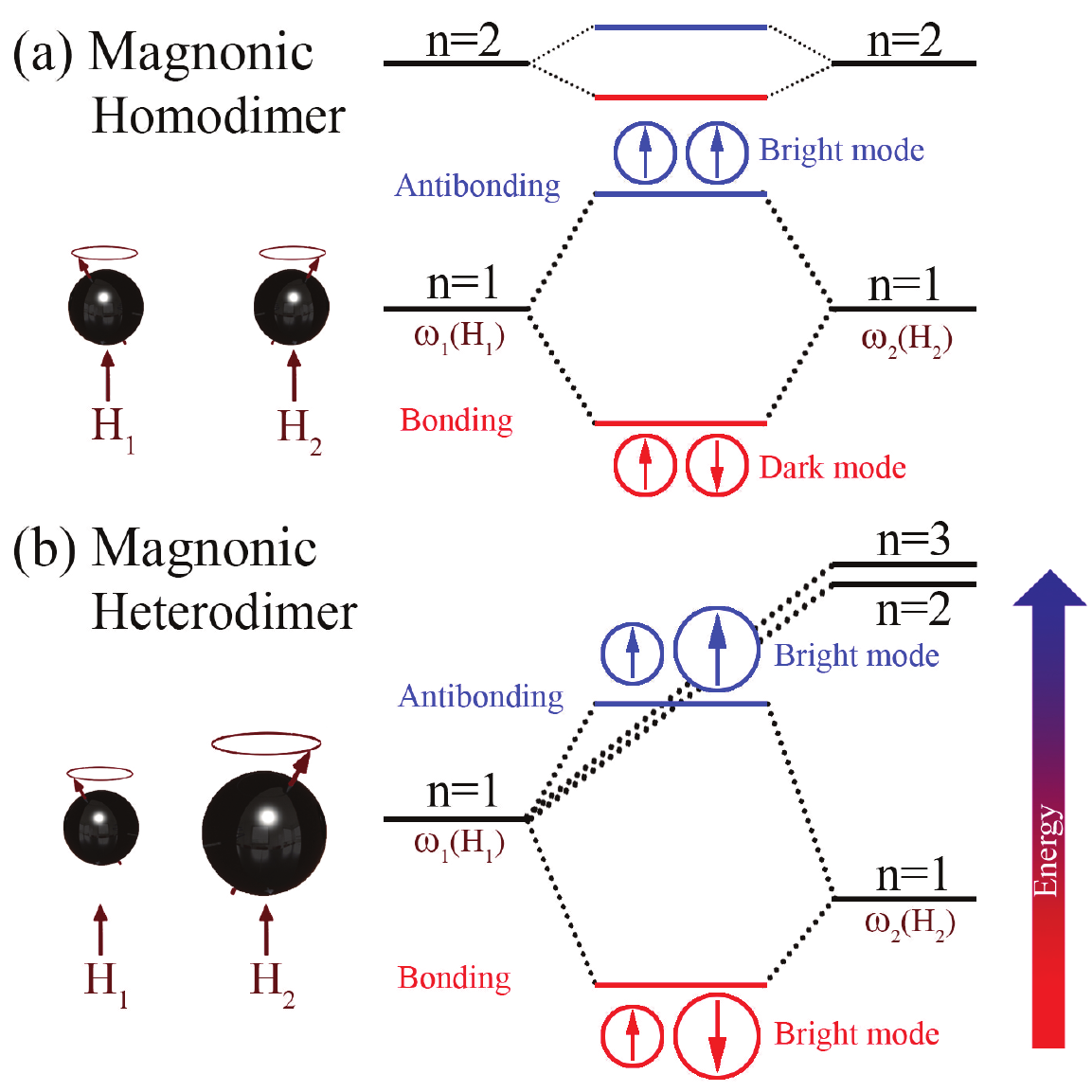}\caption{(Color online) Energy level
diagram describing the magnon hybridization in analogy with chemical bonds
resulting from the interaction between two spheres via microwave cavity modes.
(a) Magnonic homodimer consists of two similar magnetic spheres subjected to
local magnetic fields $H_{1(2)}=H_{0}\pm\delta H$, and (b) Magnonic
heterodimer consists of two dissimilar magnetic spheres. In (a) magnon
hybridization only occurs between magnonic states of the same angular momentum
denoted by $n_{i}$, while the reduced symmetry in a heterodimer introduces
coupling between all modes. In a homodimer, the bonding level is dark since it
has no dipole moment, while the antibonding level is bright. In a heterodimer
all modes are visible. The arrows in circles indicate the relative magnonic
phase (not spin or equilibrium magnetization).}%
\label{fig6}%
\end{figure}

The radiative coupling transforms the individual magnon (Kittel) modes of the
two-particle system into linear combinations, analogous to the molecular
orbital theory of diatomic molecules, according to which the interaction of
two atoms splits the levels into \textit{bonding }(symmetric) and
\textit{antibonding }(antisymmetric) orbitals. The magnetic spheres can be
interpreted as \textit{magnonic atoms} that are bound into \textit{magnonic
molecules}. Particle arrays will form \textit{magnonic crystals}, although
this term is also used for magnetic structures with periodic variations of
their magnetic properties~\cite{Krawczyk, Chumak} or distributions of
dipolar-coupled constituent materials~\cite{Vasseur}. The magnonic dimer has
bonding and antibonding combinations, where the hybridization depends on the
difference in their energies $\omega_{i}(H_{i})$ and on their interaction. A
homodimer A$_{2}$ corresponds to Fig. \ref{fig6}(a), while the mismatched
spheres in Fig. \ref{fig6}(b) form a heterodimer AB.

\begin{figure}[ptb]
\includegraphics[width=8.5cm]{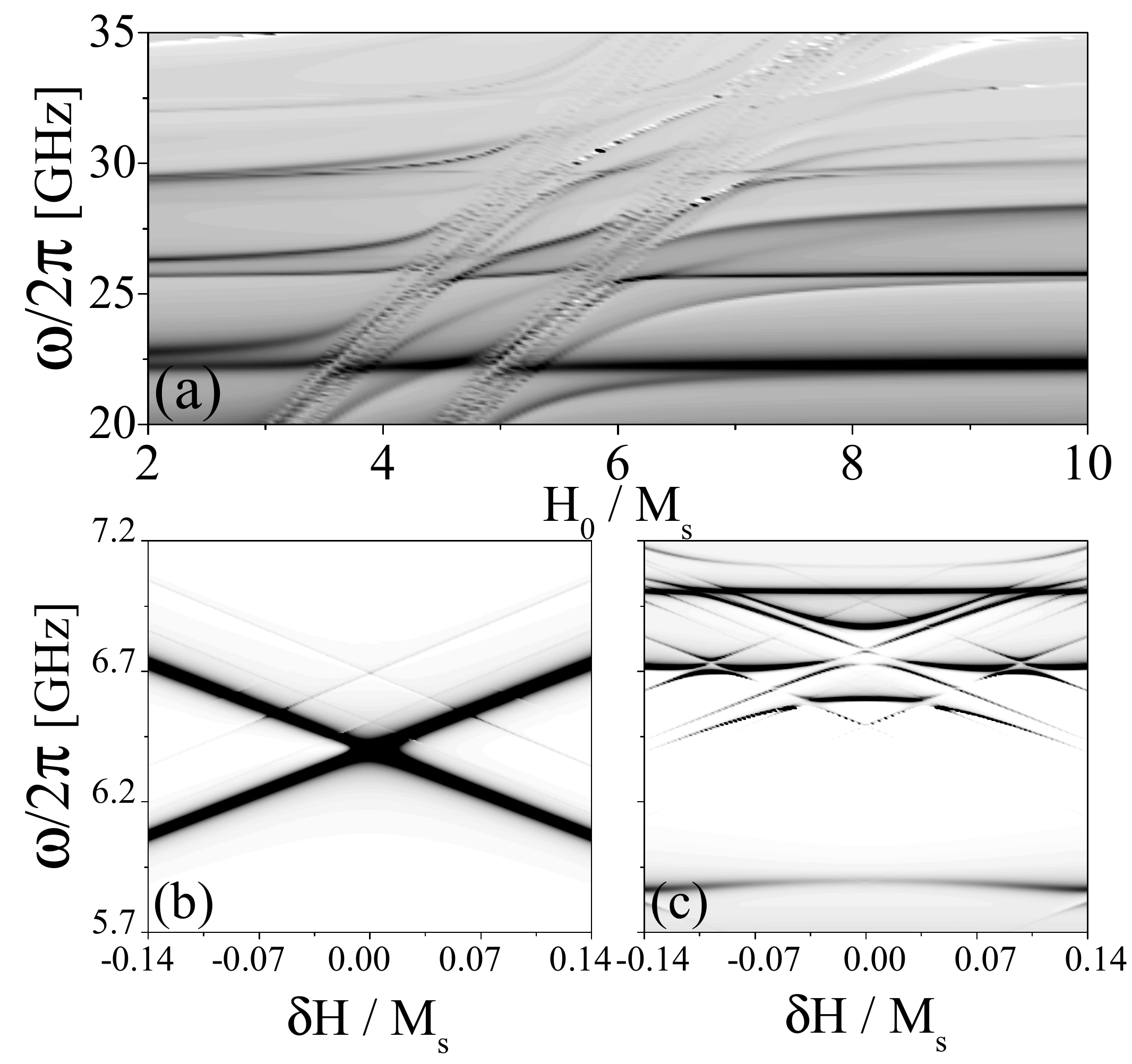}\caption{Same as Figs. \ref{fig3} and
\ref{fig4} but for relatively large YIG spheres of radius $a_{1}=a_{2}=1.25$
mm. The cavity modes are strongly mixed with those confined in the two YIG
spheres. In (a) the modes are shifted relative to each other by $\delta
H/M_{s}\sim0.7$ and the uniform field in panels (b) and (c) is fixed at
$H_{0}/M_{s}\sim6$.}%
\label{fig7}%
\end{figure}

In a homodimer with inversion symmetry in which the splitting between internal
modes is large, bonding is dominated by magnons with the same angular momentum
$n$. We may use chemical intuition, however, to maximize the coupling by
varying both the local field and the sphere radius. This may reduces the
splitting between the internal $n=1$ and $n=2$ modes (cf. Fig. \ref{fig7}) and
facilitate an increased bonding via sp-hybrid states.

Bonding and antibonding modes belong to different irreducible representations.
In a heterodimer the lack of a mirror plane reduces the spatial symmetry and
introduces couplings between all modes. Furthermore, energies of the different
shells shift with respect to each other. Fig. \ref{fig6} illustrates that the
lowest-energy (dipolar) magnon of the smaller particle can couple efficiently
to both the dipolar and higher multipolar magnons of the larger particle. The
heterodimer thereby displays a significantly more complex magnon mixing
behavior than the homodimer.

The bonding configuration corresponds to two dipole moments moving out of
phase (optical mode, negative parity of dipole moments, or antisymmetric
magnetic fields), while the antibonding configuration corresponds to the
positive parity of the dipoles (acoustic mode, symmetric fields). In contrast
to the positive parity (symmetric) magnons, the net magnetic moment of the
negative parity (antisymmetric field) magnon vanishes for identical spheres,
and does not interact with the $p$-wave cavity mode in the present
configuration. The former are then \textit{bright}, and the latter the
\textit{dark }states, as shown in Fig. (\ref{fig4}). In the heterodimer, all
magnons mix and contribute to the bonding and antibonding modes. As a
consequence, all modes become bright, see Fig. (\ref{fig5}).

We can parameterize the observations by elementary molecular orbital theory.
The energy gap, $E_{\mathrm{gap}}$, between the bonding and antibonding energy
levels for a diatomic molecule is given by the secular equation
\begin{equation}
E_{\mathrm{gap}}^{2}=(2g)^{2}+\left(  E_{A}-E_{B}\right)  ^{2}%
\end{equation}
where $g$ is the coupling parameter between the two sites, while $E_{A}$ and
$E_{B}$ refer to their energies. There are two contributions to the energy
gap, the covalent(homopolar) bonding contribution $E_{h}=2g$, and the ionic
contribution, $E_{i}=E_{A}-E_{B}$, due to the difference in \textquotedblleft
electronegativity\textquotedblright\ between the two atoms. For any bond, we
can then define the bond covalency, $\alpha_{c}=E_{h}/E_{\mathrm{gap}}$, and
polarity, $\alpha_{p}=E_{i}/E_{\mathrm{gap}}$, which parametrizes the
continuous transition from covalent to ionic bonding.

In a homodimer at $\delta H=0$ we have a direct gap due to covalent bonding
$E_{\mathrm{gap}}=2g$, see Fig. \ref{fig4}(b) and bonding and anti-bonding
wave functions are equally shared between the two atoms. However, in a
heterodimer, due to the detuning of the atomic levels $\omega_{1}(H_{0}%
)\neq\omega_{2}(H_{0})$, the gap has an \textquotedblleft
ionic\textquotedblright\ contribution, leading to an indirect gap as a
function of $\delta H$ in Fig. \ref{fig5}(b). In a polar molecule, the
amplitude of the bonding state shifts towards the more
\textit{magnon-negative} site referred to as the \textit{magnonic anion}, with
the anti-bonding state shifting towards the less magnon-negative site,
referred to as the \textit{magnonic cation}, a partially polarized molecule.
The covalent bonding strength can be independently modulated by the average
frequency spacing with the dominant cavity mode.

The scattering efficiency factor $Q_{\mathrm{sca}}$ is plotted as a function
of frequency $\omega/2\pi$, uniform field $H_{0}/M_{s}$, and differential
field $\delta H/M_{s}$ for two YIG spheres of radius $a_{1}=a_{2}=1.25$ mm and
relative permittivity $\epsilon_{\mathrm{sp}}/\epsilon_{0}=15,$ placed in a
spherical cavity of radius $R=4$ mm in Fig. \ref{fig7}(a) and (c), and without
cavity in \ref{fig7}(b). Without cavity the system can be interpreted as two
independent antennas operating in the ultrastrong coupling regime, since due
to their relatively large size individual spheres act as efficient microwave
antennas. Many anticrossings in Fig. \ref{fig7}(a) emphasize that the cavity
modes are strongly and even ultrastrongly mixed with the modes in each
individual spheres when detuned by a differential field $\delta H/M_{s}%
\sim0.7$. The large differences between Figs. \ref{fig7}(b) and (c) provide
more evidence for the strong cavity-mode induced coupling between the spheres.
In Fig. (\ref{fig7})(b), beside the main crossing modes in absence of the
cavity, we observe tails from other crossings modes at higher frequencies,
which are standing electromagnetic resonance modes confined by the magnetic
spheres. Strong coupling with cavity mode not only turns the main crossing
modes into anticrossing but also causes the complex anticrossing pattern shown
in Fig. (\ref{fig7})(c) by hybridizing all higher modes.

\section{Conclusion}

\label{sec:concl}

In conclusion, we studied the plasmonics and optomagnonics of two dielectric
and two magnetic spheres in microwave cavities by Mie scattering theory, i.e.
a systematic expansion of the coupled Maxwell and LLG equations for magnetic
systems. We employ the linear and magnetostatic approximations, but otherwise
the treatment is numerically exact. The magnetization dynamics of spatially
separated spheres in cavities can be efficiently coupled over large distances.
The main reason is not the magnetic but the electric-field coupling, since two
dielectric spheres with zero magnetization in a cavity display very similar
dynamic behavior. Both strong and ultrastrong coupling can be realized not
only for individual spheres but also for their mutual interaction. Two
(properly placed) identical spheres form an inversion symmetric system, which
is apparent by an anticrossing that generates a \textquotedblleft
direct\textquotedblright\ gap when plotted as a function of a symmetry
breaking parameter, such as a staggered magnetic field or a size difference.
Spheres with different sizes, however, break the symmetry at constant magnetic
field and lead to an \textquotedblleft indirect gap\textquotedblright\ as a
function of field detuning. Magnon-polaritons within individual magnetic
spheres may also hybridize in cavities, forming a complex mixed state of light
and spin. Our study suggests a new direction for \textquotedblleft spin
cavitronics\textquotedblright, viz. a route towards coherent control of the
dynamics of various systems and materials (magnets, pieozoelectrics,
superconductors, charge density waves, etc.) in microwave cavities via the
non-magnetic (plasmonic) interactions.

\acknowledgments B. Z. R. thanks S. M. Reza Taheri, A. Eskandari-asl, M. F.
Miri and Y. M. Blanter for fruitful discussions. This work was partially
supported by Iran Science Elites Federation (B. Z. R). Our research was
supported by the Dutch NWO and JSPS Grants-in-Aid for Scientific Research
(Grant Nos. 25247056, 25220910, 26103006).

\end{document}